\documentclass[3p,times,twocolumn]{elsarticle}

\usepackage{ecrc}

\volume{00}

\firstpage{1}

\journalname{Nuclear Physics B Proceedings Supplement}

\runauth{}

\jid{nuphbp}

\jnltitlelogo{Nuclear Physics B Proceedings Supplement}

\usepackage{amssymb}

\usepackage[utf8]{inputenc}
\usepackage{xspace}
\usepackage{hyperref}

\newcommand{\sarah}{SARAH\@\xspace}
\newcommand{\fs}{FlexibleSUSY\@\xspace}

\newcommand{\ESSM}{E$_6$SSM\@\xspace}
\newcommand{\code}[1]{\lstinline|#1|}  
\newcommand{\textoverline}[1]{$\overline{\mbox{#1}}$}
\newcommand{\DRbar}{\textoverline{DR}\xspace}

\begin{document}

\begin{frontmatter}



\dochead{}

\title{FlexibleSUSY --- a \emph{meta} spectrum generator for supersymmetric models}


\author[monash]{Peter~Athron}
\author[valencia]{Jae-hyeon~Park\corref{presenter}}
\author[dresden]{Dominik~Stöckinger}
\author[desy]{Alexander~Voigt}

\cortext[presenter]{Presenter at ICHEP 2014.}

\address[monash]{ARC Centre of Excellence for Particle Physics at the
Terascale, School of Physics, Monash University, Melbourne VIC 3800,
Australia}
\address[valencia]{Departament de Física Teòrica and IFIC,
Universitat de València-CSIC,
46100, Burjassot, Spain}
\address[dresden]{Institut für Kern- und Teilchenphysik,
TU Dresden, Zellescher Weg 19, 01069 Dresden, Germany}
\address[desy]{Deutsches Elektronen-Synchrotron DESY,
  Notkestraße 85, 22607 Hamburg, Germany}

\begin{abstract}
  \fs is a software package that takes as input descriptions of
  (non-)minimal supersymmetric models written in Wolfram/Mathematica
  and generates a set of
  spectrum generator libraries and executables, with the aid of \sarah.
  The design goals are precision, reliability, modularity, speed,
  and readability of the code.
  The boundary conditions are independent C++ objects
  that are plugged into the boundary value problem solver
  together with the model objects.
  This clean separation makes it easy to adapt
  the generated code for individual projects.
  The current status of the interface and implementation
  is sketched.
\end{abstract}


\end{frontmatter}




In a generic supersymmetric extension of the Standard Model (SM),
there are many couplings and mass parameters that can be sources of
flavour-changing neutral currents and/or $CP$-violation in addition to
the standard Cabibbo-Kobayashi-Maskawa quark mixing matrix.
Unless controlled in a particular way,
those extra sources
might cause excesses in flavour/$CP$-violating observables
that are large enough easily to kill the theory.
A popular strategy to circumvent this phenomenological danger
is to view the supersymmetric standard model (SSM)
as an effective field theory (EFT) imagining that
an underlying theory has given rise to
a reasonable pattern of the parameter values
via a certain mechanism that mediates supersymmetry breaking
in a hidden sector to the visible world.
A typical example of such a pattern consists of
universal scalar masses and
trilinear couplings proportional to the corresponding Yukawa couplings
as well as real gaugino and supersymmetric masses.  

It is unknown at which scale the underlying physics decouples
giving way to a SSM\@.
One could only speculate based on circumstantial evidences
such as gauge coupling unification which suggests that
the reasonable parameter pattern emerges at or above
the unification scale.
Neither is it clear how many stages of decoupling there are.
It is not unreasonable to consider e.g.\
a sequence of transitions like
superstring theory $\rightarrow$ supergravity $\rightarrow$
supersymmetric grand unified theory (GUT) $\rightarrow$ SSM\@.
In a more complicated scenario, each of these stages might
further consist of multiple EFTs.
A prime example would be to insert a right-handed neutrino threshold
(or multiple thresholds depending on the mass hierarchy)
between the GUT scale and the low scale.
Other variations enjoying growing interests
due to the ever rising lower bounds on the sparticle masses
from the Large Hadron Collider (LHC),
are high scale supersymmetry and split supersymmetry
which would be best described by
additional non-supersymmetric EFTs
laid below the SSM\@.

Within the above chain, 
the large scale differences among the thresholds
naturally call for treatment of
the renormalization group (RG) evolution of the parameters.
In this kind of setup,
the primary task of a spectrum generator (SG) is to
find the low-energy particle spectrum
that meets a set of boundary conditions (BCs).
Most typical BCs are
those imposed on the parameters at a high scale, $M_X$,
enforcing the aforementioned reasonable pattern.
In addition,
one requires successful electroweak symmetry breaking (EWSB)
as well as matching with the low-energy data, i.e.\ 
$\alpha_\mathrm{em}$, $\alpha_s$, $M_Z$, $M_W$,
quark and lepton masses.
If there are multiple EFTs between $M_X$ and the low scale,
one further imposes a matching condition between
each pair of adjacent EFTs.
The SG is then to solve the boundary value problem (BVP),
consisting of the RG equations (RGEs)
and the BCs at multiple scales.
After finding a solution, 
the SG calculates the pole masses and mixing
of the particles.

There are already several publicly available SGs.
Most of them have been developed for the minimal SSM (MSSM)\@.
The second best supported model is the next-to-MSSM (NMSSM)\@.
For the other models, no dedicated SG has been published
although there are codes in private use.
This is an unsatisfactory status given the wide variety of
supersymmetric models.
As the sparticle searches at the LHC keep ruling out the bulk of
the ``natural'' parameter volume of the MSSM,
studies of non-minimal SSMs are becoming more and more important.
This motivation is reinforced by the measurement of the Higgs boson mass,
which is rather high for ``natural'' accommodation within the minimal model.

\fs \cite{Athron:2014yba}, a spectrum generator generator,
is the result from
an attempt to extrapolate this narrow spectrum of spectrum generators.
It is a system of codes which accepts the description of a model
and generates a SG tailored for it.
Its major components are:
GNU-style configuration and build system,
\fs meta-code written in Wolfram/Mathematica,
\fs core library,
SG code templates,
predefined model files,
bundled codes from external sources,
programming examples and documentation.
It is intended to be
precise, reliable, modular, easy to build on, fast, capable of EFT towers,
and open to alternative BVP solving methods.
The generated SG code is written in C++,
one of the most popular programming languages.
This should make the SG code comfortably accessible
to a large number of postdocs and postgraduates.
Being an object-oriented language,
C++ allows one naturally to organize the code
in a clean modular structure that is designed directly after
the physical picture in a postdoc's mind.



\fs is not the first SG generator.
There has been already a Mathemtica package called \sarah
\cite{Staub:2012pb,Staub:2013tta},
which can also generate a SG
for a user-defined model.  
It should be worthwhile to clarify
major differences and relationship between \fs and \sarah.
\sarah generates the SG code in FORTRAN\@.
A \sarah-generated SG is linked against SPheno, whereas
a \fs-generated SG includes parts of Softsusy \cite{Allanach:2001kg}.
\sarah can produce useful physics objects for the given model such as
the interaction vertices,
$\beta$-functions, self-energies, and tadpoles, as well as
the EWSB equations.
\fs launches \sarah to obtain these results
which \fs processes subsequently to generate a C++ class library for the model.

For this reason, \fs combines two pieces of input for each model:
an input file to it, and
a set of model files to \sarah.
The \sarah model files specify
the superfield content, superpotential, gauge symmetries, and field mixing.
The \fs input file contains the rest of the instructions
to the meta-code on how to generate the SG\@.
They can be roughly grouped into two categories:
BCs on the model parameters,
and switches that control the behaviours of the SG\@.

The model definitions bundled in the package are:
the MSSM,
the $Z_3$-symmetric and the $Z_3$-violating NMSSM,
the USSM,
the non-universal-Higgs-mass (NUHM) \ESSM,
the right-handed neutrino extended MSSM,
the NUHM-MSSM,
and the $R$-symmetric MSSM\@.

A typical workflow of a postdoc/postgraduate using \fs would be as follows:
(1) install \fs and \sarah;
(2) pick out a model specification from the predefined collection or
prepare one if not already available;
(3) create a directory for the model under the \fs root;
(4) tune the \fs input file as needed;
(5) configure the build process;
(6) run \texttt{make} to build the SG executable;
(7) play with the SG,
varying the input parameters through the SUSY Les Houches Accord (SLHA)
file fed to it;
(8) pass the SLHA output on to other physics analysis tools;
(9) exploit the generated C++ class library for a more advanced study.

To formulate a new BC,
one would need to introduce a different set of input parameters.
For instance, the constrained MSSM (CMSSM) BC can be expressed
in terms of $\{m_0, m_{1/2}, \tan \beta, \mathrm{sign}(\mu), A_0\}$,
which the \texttt{MINPAR} block accommodates in the SLHA format.
In addition to these,
a NUHM BC depends on two more parameters which determine
the high-scale soft Higgs masses.
If one declares these extra parameters in the model file,
\fs automatically takes care of them and generates a C++ code
which understands their settings listed in the \texttt{EXTPAR} block
in an SLHA input file.


Based on the input files,
the \fs meta-code produces a model-specific C++ class library including
the physics building blocks such as
interaction vertices,
two-loop $\beta$-functions
with the full flavour structure of couplings and masses retained,
self-energies, and tadpoles, as well as
the procedures that calculate
the \DRbar and the pole masses and mixing.
For a generic model,
the pole masses include one-loop corrections.
For a couple of popular models, the MSSM and the NMSSM,
one can opt to add leading two-loop corrections to
the neutral Higgs pole masses, by setting the corresponding switches
in the input file.
In addition,
the library contains the BC classes as described
in the input file.
The low-energy BC class implements
the standard one-loop threshold corrections to
the gauge and the third-family fermion Yukawa couplings,
plus the two-loop QCD correction to the top quark Yukawa.
The EWSB class includes the one-loop tadpoles by default.
For the (N)MSSM,
one can choose to add the two-loop tadpole corrections.
The driver template is then instantiated
to the given model which forms the \texttt{main()} function
of the SG program.
Finally, \fs links the driver and the model library against the
core library and external dependencies
to produce the SG executable.


The operation of the program is like other SGs:
(1) read the input parameters from a file;  
(2) make an initial guess of the model parameters;
(3) enter the fixed-point iteration loop in which
the low-scale, the high-scale, and the SUSY-scale BCs
are enforced in the cyclic order after bringing
the model parameters to each scale by integrating the RGEs;
(4) after the model parameters converge to a fixed-point,
compute the pole mass spectrum;
(5) produce the output.  

The model-specific $\beta$-functions and the BCs,
implemented as C++ classes, belong to each model library.
In a SG program,
objects of these RGEs and BCs are plugged into
the BVP solver object whose implementation is located in the core library.
In fact, this solver is already ready to accept multiple models,
even though the current version of \fs
auto-generates a SG only for a single model configuration.
Given hand-written matching condition classes,
it is straightforward to author a SG for a tower of EFTs
as mentioned in the early part of the article.
The \fs package comes with such an example which realizes
the type-I see-saw mechanism by stacking
the MSSM plus three heavy right-handed neutrinos
on top of the MSSM including the dimension-5 operator.


We take great care in the validation of \fs.
Various stages of its operation
have been and are being tested thoroughly against
Softsusy \cite{Allanach:2001kg}
and Next-to-Minimal Softsusy \cite{Allanach:2013kza}.
The tested components include:
the (N)MSSM $\beta$-functions,
self-energies and tadpoles,
tree-level masses and mixing,
pole masses,
EWSB conditions at the tree-level and loop-levels,
as well as the core routines such as
the Runge-Kutta differential equation integrator,
the Passarino-Veltman functions,
and the linear algebra procedures.
An automatic system carries out an extensive suite of tests every night
on the source code snapshot fetched from the repository hosted at github.com.
Thanks to these regular tests,
there has been a notable case where we could spot a Heisenbug
due to a race condition caused by a thread-unsafe subroutine
for evaluating two-loop Higgs mass corrections.

For an efficient research activity of a human being,
the responsiveness of a system is an important factor.
Even a delay of half a minute
between the input action and the output could easily
lead the researcher to a distraction such as web surfing.
Fast execution per each input point would also greatly aid
a scan or global fit in multidimensional parameter space.
In this respect, \fs aims to be a competitive choice.
With the flavour-off-diagonal sfermion self-energies ignored
in the pole mass calculation,
\fs is faster by a factor of 1.4--1.7 than SPheno 3.2.4,
and by a factor of 2--2.5 than Softsusy 3.4.0.
With the full $6 \times 6$ sfermion mass matrices
taken into account,
\fs is faster by a factor of 2.8--5 than
a \sarah-generated SPheno-like SG.
\fs can be further accelerated on a platform with multiple CPU cores
thanks to its multi-threaded pole mass computation.



There are already exciting studies
making use of \fs \cite{Diessner:2014ksa,Athron:2014pua}.
More features are coming.
Stay tuned.


J.P. acknowledges support from the MEC and FEDER (EC) Grants
FPA2011--23596 and the Generalitat Valenciana under grant PROMETEOII/2013/017.





\begin{thebibliography}{0}



\bibitem{Athron:2014yba}
P.~Athron, J.-h. Park, D.~Stöckinger, A.~Voigt, FlexibleSUSY --- a spectrum
  generator generator for supersymmetric models.
\newblock \href
  {http://arxiv.org/abs/1406.2319} {\path{arXiv:1406.2319}}.

\bibitem{Staub:2012pb}
F.~Staub, {SARAH 3.2: Dirac Gauginos, UFO output, and more}, Computer Physics
  Communications 184 (2013) pp. 1792--1809.
\newblock \href {http://arxiv.org/abs/1207.0906} {\path{arXiv:1207.0906}},
  \href {http://dx.doi.org/10.1016/j.cpc.2013.02.019}
  {\path{doi:10.1016/j.cpc.2013.02.019}}.

\bibitem{Staub:2013tta}
F.~Staub, {SARAH 4: A tool for (not only SUSY) model builders},
  Comput.Phys.Commun. 185 (2014) 1773--1790.
\newblock \href {http://arxiv.org/abs/1309.7223} {\path{arXiv:1309.7223}},
  \href {http://dx.doi.org/10.1016/j.cpc.2014.02.018}
  {\path{doi:10.1016/j.cpc.2014.02.018}}.

\bibitem{Allanach:2001kg}
B.~Allanach, {SOFTSUSY: a program for calculating supersymmetric spectra},
  Comput.Phys.Commun. 143 (2002) 305--331.
\newblock \href {http://arxiv.org/abs/hep-ph/0104145}
  {\path{arXiv:hep-ph/0104145}}, \href
  {http://dx.doi.org/10.1016/S0010-4655(01)00460-X}
  {\path{doi:10.1016/S0010-4655(01)00460-X}}.

\bibitem{Allanach:2013kza}
B.~Allanach, P.~Athron, L.~C. Tunstall, A.~Voigt, A.~Williams, {Next-to-Minimal
  SOFTSUSY}, Comput.Phys.Commun. 185 (2014) 2322--2339.
\newblock \href {http://arxiv.org/abs/1311.7659} {\path{arXiv:1311.7659}},
  \href {http://dx.doi.org/10.1016/j.cpc.2014.04.015}
  {\path{doi:10.1016/j.cpc.2014.04.015}}.

\bibitem{Diessner:2014ksa}
P.~Dießner, J.~Kalinowski, W.~Kotlarski, D.~Stöckinger, {Higgs boson mass and
  electroweak observables in the MRSSM}.
\newblock \href {http://arxiv.org/abs/1410.4791}
  {\path{arXiv:1410.4791}}.

\bibitem{Athron:2014pua}
P.~Athron, M.~Muhlleitner, R.~Nevzorov, A.~Williams, {Non-Standard Higgs Decays
  in $U(1)$ Extensions of the MSSM}.
\newblock \href {http://arxiv.org/abs/1410.6288}
  {\path{arXiv:1410.6288}}.

\end{thebibliography}



\end{document}